\documentclass[12pt]{article}
\usepackage{graphicx}
\usepackage{amssymb}
\usepackage{amscd}
\usepackage{amsmath}

\begin{document}
\begin{titlepage}

\vskip 1cm
{\hbox to\hsize{\hfill September 2014 }}

\vskip 0.6cm

\bigskip \vspace{3\baselineskip}

\begin{center}
{\bf \large 

Criticality in the scale invariant standard model (squared)  }

\bigskip

\bigskip

\vskip 1cm

{\bf Robert Foot$^{\rm a}$, Archil Kobakhidze$^{\rm b}$ and Alexander Spencer-Smith$^{\rm b}$ \\ }

\smallskip
\vskip 0.6cm

{ \small \it
ARC Centre of Excellence for Particle Physics at the Terascale, \\
$^{\rm a}$School of Physics, The University of Melbourne, VIC 3010, Australia \\
$^{\rm b}$School of Physics, The University of Sydney, NSW 2006, Australia \\
E-mails: rfoot@unimelb.edu.au, archilk@physics.usyd.edu.au, alexss@physics.usyd.edu.au 
\\}

\bigskip
 
\bigskip

\bigskip
\vskip 1cm

{\large \bf Abstract}

\end{center}
\noindent 
We consider first the standard model Lagrangian with $\mu_h^2$ Higgs potential term set to zero.
We point out that this clasically scale invariant theory potentially exhibits radiative electroweak/scale symmetry breaking 
with very high vacuum expectation value (VEV) for the Higgs field, $\langle \phi \rangle \approx 10^{17-18}$ GeV.
Furthermore, if such a vacuum were realized then cancellation of vacuum energy automatically implies that this nontrivial vacuum is 
degenerate with the trivial unbroken vacuum. Such a theory would therefore be critical with the Higgs self-coupling and its beta 
function nearly vanishing at the symmetry breaking minimum, $\lambda (\mu=\langle \phi \rangle)\approx \beta_{\lambda} (\mu=\langle \phi \rangle)\approx 0$.  
A phenomenologically viable model that predicts this criticality property arises
if we consider two copies of the standard model Lagrangian, with exact $Z_2$ symmetry swapping each ordinary
particle with a partner.  The spontaneously broken vacuum can then arise where one sector gains the high scale VEV, 
while the other gains the electroweak scale VEV. 
The low scale VEV is perturbed away from zero
due to a Higgs portal coupling, or via the usual small Higgs mass terms $\mu_h^2$, which softly break the scale invariance.
In either case, the cancellation of vacuum energy requires $M_t = (171.53 \pm 0.42)$ GeV,  which is close
to its measured value of $(173.34 \pm 0.76)$ GeV. 

\end{titlepage}


The discovery of a Higgs-like particle \cite{exp1,exp2} with mass around
125 GeV confirms the standard picture of electroweak symmetry breaking 
via the nonzero vacuum expectation value of a scalar field \cite{higgs}. 
This nontrival vacuum arises from a Higgs potential, of the form:
\begin{eqnarray}
V = \lambda \phi^{\dagger}\phi \phi^{\dagger}\phi - \mu_h^2
\phi^{\dagger}\phi + h\mu_h^4 \ . 
\end{eqnarray}
where the $h\mu_h^4$ part is the 
cosmological constant (CC) term, usually neglected as it only affects gravitational physics, e.g. \cite{jap,jones}. 
In models with classical scale invariance, however, the
CC term is absent, as required by this symmetry.
The physical cosmological constant still arises radiatively, but is a calculable function of
the other parameters of the theory \cite{Foot:2010et}.

Interestingly, a possible hint of a deeper structure beyond the standard
model has emerged in a rather
unexpected manner. The Higgs quartic coupling when evolved up to a high
scale $\sim 10^{17-18}$ GeV, appears to approximately satisfy: $\lambda =
\dot{\lambda} (\equiv \beta_{\lambda}) = 0$ (for recent calculations, see \cite{buz,alex}).
This condition seems to be accidental, since it involves cancellation among numerically large 
quantitites\footnote{Previous attempts to justify these conditions were based on somewhat obscure principle of 
multiple criticality \cite{Froggatt:1995rt} and still controversal proposal of 
asymptotic safety of gravity \cite{Shaposhnikov:2009pv}. See also \cite{Gorsky:2014una}.}. 
In this short note, we show that such a relation can naturally arise in scale
invariant models as a consequence of setting the physical cosmological constant to 
its measured small value. This then automatically implies that two distinct vacua with broken 
and unbroken symmetries coexists in the theory, that is, the theory exhibits criticality.  
We consider cases of exact classical scale invariance
and also the case where scale invariance is considered to be softly
broken by the familiar $\mu_h^2$ term in the Higgs potential.

Let us define the scale invariant standard model Lagrangian, ${\cal L}_{SM}^{SI}$, to be
the same as the standard model Lagrangian except with the $\mu_h^2$ term set to zero.
The Higgs potential is then particularly simple:
\begin{eqnarray}
V = \lambda \phi^{\dagger}\phi \phi^{\dagger} \phi
\ .
\end{eqnarray}
A Coleman-Weinberg analysis \cite{Coleman:1973jx,GildenerWeinberg},
reveals that such a potential, radiatively corrected, can exhibit
spontaneous symmetry breaking. This requires $\lambda (\mu)$ to be small
at some particular scale $\mu = \mu_1$, and $\mu_1$ sets the scale of
the VEV of $\phi$. If we additionally require that the CC vanish, then
we have much more stringent
constraints for spontaneous symmetry breaking to occur
\cite{Foot:2010et}. To explain what these are, let us first 
write the exact (all loop) effective potential as:
\begin{eqnarray}
V=A(g_a(\mu), m_{x}(\mu),\mu) \phi^{\dagger}\phi
\phi^{\dagger}\phi+B(g_a(\mu), m_{x}(\mu),\mu) 
 \phi^{\dagger}\phi \phi^{\dagger}\phi
\log\left(\frac{\phi^{\dagger}\phi}{\mu^2}\right) \nonumber \\
+C(g_a(\mu), m_{x}(\mu),\mu) \phi^{\dagger}\phi \phi^{\dagger}\phi
\left[\log\left(\frac{\phi^{\dagger}\phi)}{\mu^2}\right)\right]^2+\ldots~,
\label{a4}
\end{eqnarray}
where $\ldots$ denotes all terms with  higher-power logarithms and
$g_a(\mu)$ and $m_x(\mu)$ denote all relevant 
running dimensionless couplings and effective masses.
Then it can easily be shown that the requirement of a nontrivial vacuum,
$\langle \phi \rangle \neq 0$, and vanishing CC require the condition:
\begin{eqnarray}
A(g_a(\mu), m_{x}(\mu),\mu) = B(g_a(\mu), m_{x}(\mu),\mu) = 0
\end{eqnarray}
to hold at the same renormalization scale $\mu = \mu_1$. Furthermore
this renormalization scale defines the VEV of $\phi$ at that scale.
Note that the renormalization scale independence of the effective
potential implies that:
\begin{eqnarray}
B (\mu = \mu_1) &=& 
\left. {1 \over 2} \mu {d A \over d\mu}\right |_{\mu = \mu_1}\ , \nonumber
\\ 
C (\mu = \mu_1) &=& \left. {1 \over 4} \mu {d B \over
d \mu}\right |_{\mu = \mu_1}~.
\label{5}
\end{eqnarray}
The condition that $A=B=0$ at
the same renormalization scale $\mu=\mu_1$ 
implies that
$\lambda \simeq 0$ and $\dot{\lambda} \simeq 0$ at that renormalization
scale. Interestingly it has been observed that such a condition is nearly
satisfied  given the parameters of the standard model.
However, one finds that the scale, $\mu_1$ is very high, 
$\sim 10^{17-18}$ GeV, that is possibly as high as the Planck scale.
Thus, it appears that the scale invariant standard model may exhibit
spontaneous symmetry breaking, but with very high VEV scale, $\langle \phi \rangle \approx 10^{17-18}$ GeV.

This spontaneously broken phase, is degenerate with the unbroken phase
where $\phi = 0$. Could it be possible that there is a second copy of
the
standard model, i.e.
mirror model with exact $Z_2$ invariant Lagrangian \cite{Foot:1991bp},
where the $Z_2$ symmetry is spontaneously broken?
That is, where one copy gets a zero VEV and the other at a large scale,
$\mu_1 \approx 10^{17-18}$ GeV? 
A small portal coupling $\kappa \phi^{\dagger}\phi \phi'^{\dagger}\phi'$
[the particles of the copy are denoted with a prime ($'$)]
could then perturb the zero VEV of $\phi$, to be nonzero, and thus be
responsible for electroweak symmetry breaking.
This could work if $\kappa > 0$ at the high scale, $\mu_1$, and $\kappa < 0$ 
at the electroweak scale.
Since the required value of $\kappa$ is very small at the electroweak
scale ($\kappa \sim 10^{-31\pm1)}$), it is possible that quantum gravitational
corrections contribute to the running of $\kappa$ and could lead to its
change in sign at the high scale cf. with the low scale.
Such a model could also be technically natural, despite the high scale (for discussions
along these lines see 
\cite{Foot:2007iy,Foot:2013hna} and references there-in).

The scale invariant standard model squared has Lagrangian:
\begin{eqnarray}
{\cal L} = {\cal L}_{SM}^{SI} (e, u, d,\gamma, ...) + {\cal L}_{SM}^{SI} (e', u', d', \gamma',...) + {\cal L}_{int}
\ .
\end{eqnarray}
The Lagrangian ${\cal L}_{int}$ contains the Higgs portal interaction and, potentially, also gauge kinetic mixing.
The potential for this scale invariant standard model squared is just the obvious generalization of Eq.(\ref{a4}),
with $A', B', C'$ parameterizing the corresponding terms for $\phi'$. 
At the high scale, $\mu=\mu_1$, the quantum corrected effective
potential has a particularly simple form. 
Given that the extremum and CC conditions
require that $A' = B' = 0$ (at that scale), the $Z_2$ symmetry also
implies $A = B = 0, \ C'=  C$ (at that scale).
It follows that
\begin{eqnarray}
V = C \phi^{\dagger}\phi \phi^{\dagger}\phi \ \left[ \log
(\phi^{\dagger} \phi /\mu_1^2)\right]^2 
+  C \phi'^{\dagger}\phi' \phi'^{\dagger}\phi' \ \left[ \log
(\phi'^{\dagger} \phi' /\mu_1^2 )\right]^2  + \kappa
\phi'^{\dagger}\phi' \phi^{\dagger}\phi
\ .
\end{eqnarray}
The leading order contribution to $C$ arises at two loops and is
given by:
\begin{eqnarray}
C^{(2)} (\mu = \mu_1) = {1 \over 64 \pi^2 \mu_1^4}\left. \left[ 3 {\rm
Tr} m_V^4 \gamma_V + {\rm Tr}
m_S^4\gamma_S - 4 {\rm Tr} m_F^4 \gamma_F \right]\right |_{\mu =
\mu_1}~, 
\label{9}
\end{eqnarray}
where $\gamma_x = \partial \ln m_x/\partial \ln \mu$ ($x = V, S, F$).  As
discussed  in the
previous paragraph, we assume $\kappa > 0$ at the high scale
$\mu=\mu_1$.
If $C>0$ at the scale $\mu = \mu_1$, the above potential  has the
spontaneously broken vacuum $\langle \phi' \rangle = \mu_1, \ \langle
\phi \rangle = 0$ (where the VEV's are running parameters defined at the
scale $\mu=\mu_1$),
as well as a degenerate unbroken vacuum: $\langle \phi' \rangle =
\langle \phi \rangle = 0$.

The Pseudo Goldstone Boson (PGB) of scale invariance,  $m_{\rm PGB} =
\left.\frac{\partial^ 2V }{\partial \phi'^2_0}\right |_{\phi'=\mu_1}$ 
arises at the  two-loop level,    
\begin{eqnarray}
m_{\rm PGB}^2 = 4C (\mu = \mu_1) \mu_1^2~.
\label{3}
\end{eqnarray}
Considering only the dominant ($W', Z'$ and $t'$) contributions,
we have
\begin{eqnarray}
C^{(2)} &=& {1 \over 64 \pi^2 \mu_1^4} \left\{ 6M_{W}^4 \gamma_{W} +
3M_{Z}^4 \gamma_{Z} - 12M_{t}^4 \gamma_{t} \right\}
\nonumber \\
&=& {1 \over 64 \pi^2 \mu_1^4} \left\{ 6M_{W}^4 {\beta_{g_2}\over g_2} +
3M_{Z}^4 \left[\cos^2 \theta_w {\beta_{g_2} \over g_2} + \sin^2 \theta_w {\beta_{g_1}\over g_2} \right] - 12M_{t}^4 {\beta_{y_{t}}\over
y_{t}} \right\}
\end{eqnarray}
where $g_1, g_2, y_{t}$ and the $U(1)$, $SU(2)$ gauge couplings
and $t$ Yukawa coupling ($\tan\theta_w \equiv g_1/g_2$) evaluated at the high scale, $\mu = \mu_1$. 
Also, the beta functions are defined by: $\beta_{X} \equiv \partial X/\partial \ln \mu$. 
Evaluating $C^{(2)}$ at the scale $\mu = \mu_1$ we find that $C^{(2)} \approx 2 \times 10^{-6}$, and thus
the PGB mass is around $m_{\rm PGB} \approx 3\times 10^{-3} \mu_1 \sim 10^{15}$ GeV. 

At the low scale $\mu \sim 100$ GeV, a non-zero VEV of $\phi$ is induced
via the portal coupling $\kappa$, provided $\kappa < 0$ at this scale.
Indeed at this low scale the part of the potential involving $\phi$ has
the approximate form:
\begin{eqnarray}
V = \lambda \phi^{\dagger}\phi \phi^{\dagger}\phi + \kappa
\phi^{\dagger}\phi \phi'^{\dagger}\phi'   
\ .
\end{eqnarray}
We expect that $\langle \phi' \rangle$, evaluated as a function of
renormalization scale, $\mu$, 
does not greatly change in going from the high scale to the low scale.
This means that the $\phi$ part of the potential, at
the electroweak renormalization scale, is just:
\begin{eqnarray}
V = \lambda \phi^{\dagger}\phi \phi^{\dagger}\phi - \mu_h^2
\phi^{\dagger}\phi
\end{eqnarray}
where 
\begin{eqnarray}
\mu_h^2 = -\kappa \mu_1^2 = -\kappa (\langle \phi' \rangle)^2 \ .
\end{eqnarray}
The correct vacuum results provided that $\mu_h^2 $ is the standard
electroweak value.

Formally, the model has the same number of Lagrangian parameters as the
standard model (just two in higgs potential, $\lambda, \kappa$), but
since the CC is predicted, it effectively has one parameter less when
coupled to gravity.
Thus, the model makes a prediction, which can be obtained from the
condition: 
\begin{eqnarray}
A'(\mu_1) = B'(\mu_1)=0 \ .
\end{eqnarray}
Up to small threshold corrections at the high scale (small
because $M_{t'}, M_{W'}, M_{Z'}$ have masses somewhat below
$\mu_1$),
the $Z_2$ symmetry in the model then implies that
\begin{eqnarray}
A(\mu_1)=B(\mu_1) = 0 \ .
\label{buz}
\end{eqnarray}
Interestingly, this condition coincides with the criticality condition
in the standard model, which
has been studied recently in several papers \cite{buz,alex}.
These studies have found that criticality in the standard model occurs
when the top quark pole mass satisfies \cite{buz}:
\begin{eqnarray}
M_t = (171.53 \pm 0.42) \ {\rm GeV} 
\label{mon}
\end{eqnarray}
where the various errors have been added in quadrature\footnote{In light of resent BICEP2 result on primordial gravitational 
waves \cite{Ade:2014xna}, the above criticality bound is also necessary to guarantee the stability of the electroweak 
vacuum during cosmic inflation \cite{Espinosa:2007qp, Kobakhidze:2013tn, alex}.}.
This value is very
close to the measured value of $173.34 \pm 0.76$ GeV \cite{ATLAS:2014wva}.
Thus, the scale invariant standard model squared requires that
the top quark should have mass around 171 GeV [Eq.(\ref{mon})], which is only around $2\sigma$ away
from its measured value.

There is a phenomenologically equivalent theory, where $\kappa = 0$, but
scale invariance is broken softly (as well as spontaneously) by the
usual soft
scale invariant breaking term in both sectors:
\begin{eqnarray}
 V_{soft} = -\mu_h^2 \phi^{\dagger}\phi -\mu_h^2 \phi'^{\dagger}\phi' + h\mu_h^4 \ .
\label{19x}
\end{eqnarray}
In this case, there is still a radiatively induced vacuum at the high
scale, $\mu_1 \approx 10^{17-18}$ GeV,
but it is meta stable, with lifetime longer than the hubble time. The
$Z_2$ symmetry is still an exact Lagrangian symmetry,
but will be  spontaneously broken when $\phi$ has electroweak scale VEV,
and $\phi'$ gets the high scale VEV. The cosmological constant in such a
theory again is completely determined in terms of existing parameters,
with essentially the same predictions for the top quark following from
demanding that the CC by small. [There is only a very tiny
correction to the prediction $\lambda (\mu_1) = 0$ of order $\delta \lambda \sim \mu^2/\langle \phi' \rangle^2$ arising
from the soft breaking terms in Eq.(\ref{19x}).]

The result that $\langle \phi' \rangle \sim M_{Planck}$ is very interesting and suggests that this
radiatively induced scale might be the origin of the Planck scale.
This can easily be realized if $\phi'$ couples to the Ricci scalar:
\begin{eqnarray}
{\cal L} = \sqrt{-g}\ \xi (\phi^{\dagger}\phi + \phi'^{\dagger}\phi') R
\ .
\end{eqnarray}
In this context, 
we would like to point out that a recently proposed successful scenario of Higgs 
inflation \cite{Hamada:2014iga, Bezrukov:2014bra} can be realised within the model described in this paper.

One could consider a generalization of the model to $N > 2$ sectors, i.e. scale invariant $[SM]^N$ model.
A permutation symmetry, $S_N$, generalizing the $Z_2$ symmetry of the scale invariant standard
model squared model, could be assumed.
This symmetry 
would be spontaneously broken at the high scale, if one of these sectors gains the large Planck scale VEV,
with $N-1$ sectors then having electroweak symmetry breaking scale.
That is, the permutation symmetry $S_N$ 
would be spontaneously broken down to a $S_{N-1}$ subgroup. For example, nature might come with
three copies of the standard model and if one 
sector gains the large Planck scale VEV, the other two can be both broken at the electroweak scale, with an exact
unbroken $Z_2$ symmetry remaining.
The extra electroweak scale copy, in this case, can provide an interesting candidate for dark matter, e.g. \cite{Foot:2014mia}
and references there-in.

In conclusion, 
we have considered first the standard model Lagrangian with $\mu_h^2$ Higgs potential term set to zero.
This clasically scale invariant theory potentially exhibits radiative electroweak/scale symmetry breaking 
at a very high scale near the Planck scale, $\langle \phi \rangle \approx 10^{17-18}$ GeV.
Furthermore, if such a vacuum were realized then cancellation of vacuum energy automatically implies that this nontrivial vacuum is 
degenerate with the trivial unbroken vacuum. Such a theory would therefore be critical with the Higgs self-coupling and its beta 
function nearly vanishing at the symmetry breaking minimum, $\lambda (\mu=\langle \phi \rangle)\approx \beta_{\lambda} (\mu=\langle \phi \rangle)\approx 0$.  
A phenomenologically viable model that predicts this criticality property arises
if we consider two copies of the standard model Lagrangian, with exact $Z_2$ symmetry swapping each ordinary
particle with a partner.  The spontaneously broken vacuum can then arise where one sector gains the high scale VEV, 
while the other gains the electroweak scale VEV. 
The low scale VEV is perturbed away from zero
due to a Higgs portal coupling, or via the usual small Higgs mass terms $\mu_h^2$, which softly break the scale invariance.
In either case, the cancellation of vacuum energy requires $M_t = (171.53 \pm 0.42)$ GeV,  which is close
to its measured value of $(173.34 \pm 0.76)$ GeV.

\vskip 1.4cm
\noindent
{\large \bf Acknowledgements}

\vskip 0.2cm
\noindent
This work was supported by the Australian Research Council. One of the authors (R. F.)
wishes to thank Michael Schmidt and Ray Volkas for useful discussions.

\end{document}